\documentclass[preprint]{emulateapj}

\received{}
\revised{}
\accepted{}

\shortauthors{Boyer et al.}
\shorttitle{``RGB Dust Production in 47~Tuc''}

\begin{document}

\title{Is Dust Forming on the Red Giant Branch in 47\,Tuc?}

\author{Martha~L.~Boyer\altaffilmark{1},
  Jacco~Th.~van~Loon\altaffilmark{2},
  Iain~McDonald\altaffilmark{3},
  Karl~D.~Gordon\altaffilmark{1},
  Brian~Babler\altaffilmark{4},
  Miwa~Block\altaffilmark{5},
  Steve~Bracker\altaffilmark{4},
  Charles~Engelbracht\altaffilmark{5},
  Joe~Hora\altaffilmark{6},
  Remy~Indebetouw\altaffilmark{7},
  Marilyn~Meade\altaffilmark{4},
  Margaret~Meixner\altaffilmark{1},
  Karl~Misselt\altaffilmark{5},
  Marta~Sewilo\altaffilmark{1},
  Bernie~Shiao\altaffilmark{1}, and
  Barbara~Whitney\altaffilmark{8}}

  \altaffiltext{1}{STScI, 3700 San Martin Drive, Baltimore, MD 21218 USA; mboyer@stsci.edu}
  \altaffiltext{2}{Astrophysics Group, Lennard-Jones Laboratories, Keele University, Staffordshire ST5 5BG, UK}
  \altaffiltext{3}{Jodrell Bank Centre for Astrophysics, Alan Turing Building, University of Manchester, M13 9PL, UK}
  \altaffiltext{4}{Department of Astronomy, University of Wisconsin, Madison, 475 North Charter Street, Madison, WI 53706-1582 USA}
  \altaffiltext{5}{Steward Observatory, University of Arizona, 933 North Cherry Avenue, Tucson, AZ 85721 USA}
  \altaffiltext{6}{Harvard-Smithsonian Center for Astrophysics, 60 Garden Street, MS 65, Cambridge, MA 02138-1516 USA}
  \altaffiltext{7}{Department of Astronomy, University of Virginia, P.O. Box 3818, Charlottesville, VA 22903-0818 USA}
  \altaffiltext{8}{Space Science Institute, 4750 Walnut Street, Suite 205, Boulder, CO 80301 USA}

\begin{abstract}
Using {\it Spitzer} IRAC observations from the SAGE-SMC Legacy program
and archived {\it Spitzer} IRAC data, we investigate dust production
in 47\,Tuc, a nearby massive Galactic globular cluster. A previous
study detected infrared excess, indicative of circumstellar dust, in a
large population of stars in 47\,Tuc, spanning the entire Red Giant
Branch (RGB). We show that those results suffered from effects caused
by stellar blending and imaging artifacts and that it is likely that
no stars below $\approx$1~mag from the tip of the RGB are producing
dust. The only stars that appear to harbor dust are variable stars,
which are also the coolest and most luminous stars in the cluster.
\end{abstract}


\section{INTRODUCTION}
\label{sec:intro}

Stellar mass loss is critical to the late stages of stellar
evolution. Mass loss in Red Giant Branch (RGB) stars determines
cluster Horizontal Branch (HB) morphology
\citep[e.g.,][]{rood73,catelan00} and the pulsational properties of RR
Lyrae variables \citep[e.g.,][]{christy66,caloi08}. Strong mass loss
on the Asymptotic Giant Branch (AGB) can exceed the nuclear
consumption rate, determining the timescale of subsequent evolution
\citep[e.g.,][]{vanloon99,boyer09b}. AGB winds are enriched due to a
convective process that brings processed material to the stellar
surface \citep[the third dredge-up; e.g.,][]{renzini81}, making AGB
stars a major source of galaxy enrichment \citep{gehrz89}. In the
low-mass stars of old globular clusters (GCs; $M < 0.8~M_\odot$), more
mass is lost on the RGB than the AGB \citep[cf.][]{mcdonald09}.

Mass loss can arise from acoustic and/or electro-magnetic
chromospheric activity \citep{hartmann80,hartmann84,vanloon10}, or
from pulsationally-enhanced, dust-driven winds
\citep[e.g.,][]{bowen88, vanloon08b}.  Radial pulsations levitate
material to radii where the stellar atmosphere is cool, and
pulsational shocks provide the density required for dust
condensation. Radiation pressure on the grains and dust-gas momentum
coupling drive the wind. However, it is uncertain whether the dust is
the driving mechanism or is simply formed as a byproduct of the wind
\citep{vanloon05,ferrarotti06,hofner07,hofner08}.

With the launch of the {\it Spitzer Space Telescope} \citep{werner04},
circumstellar dust has become easily detectable, leading to a more
complete picture of dust-driven mass loss: (1) dust forms only at or
above the tip of the RGB (TRGB) \citep[e.g.,][]{boyer08,boyer09a}; (2)
dust formation is not inhibited at low metallicity
\citep[e.g.,][]{boyer06,boyer09b,sloan09}; and (3) dust formation
appears episodic, possibly tied to the pulsation cycle
\citep[e.g.,][]{origlia02,mcdonald09}.  The first of these points has
been more controversial. \citet{kalirai07} speculate that strong mass
loss in RGB stars might explain the presence of He white dwarfs
in NGC~6791, although \citet{vanloon08a} find that, along with an
absence of dust, the RGB and HB luminosity function is not
depleted. \citet{origlia07} find 93 stars with infrared (IR) excess as
faint as the HB in 47\,Tuc, and \citet{boyer06} find 12 such sources
in M15 (see their Fig.~3b). The presence of dusty stars so far down
the RGB is surprising given low luminosities, warm photospheres,
and a lack of pulsation. Indeed, more recent {\it Spitzer} observations
show very few dusty stars in NGC~362 \citep{boyer09a} and
$\omega$\,Cen \citep{boyer08,mcdonald09}.

This letter examines dust production in 47\,Tuc, which is among the
most massive \citep[$M_V = -9.42$~mag\footnote{The \citet{harris96}
catalog was updated in 2003:
http://www.physics.mcmaster.ca/$\sim$harris/mwgc.dat};][]{harris96},
metal-rich \citep[${\rm [Fe/H]} = -0.7$;][]{thompson09} and nearby GCs
\citep[$(m-M)_0 = 13.32$;][]{ferraro99}. We present new {\it Spitzer}
observations along with a re-analysis of the \citet{origlia07}
data. We discuss how stellar blending affects the low-resolution
8~\micron{} photometry and find that there is little, if any, dust
below the TRGB in 47\,Tuc.

\begin{figure}
\epsscale{1.15} \plotone{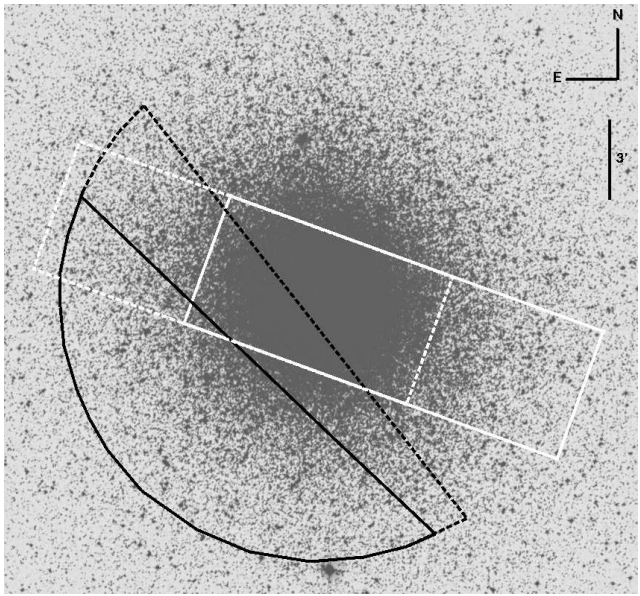} \figcaption{DSS image of 47\,Tuc showing
the {\it Spitzer} coverage. The SAGE-SMC coverage used here is shown
in black and the Rood coverage in white.  The 3.6 and 8~\micron{}
coverages are represented by dashed and solid lines, respectively. We
limited the SAGE-SMC coverage to $R<10\arcmin$ to minimize
contamination from the SMC and to mimic the Rood coverage.
\label{fig:img} }
\end{figure}

\section{Data and Analysis}
\label{sec:data}

The small separation ($\approx$2\degr) between 47\,Tuc and the Small
Magellanic Cloud (SMC) resulted in serendipitous {\it Spitzer}
observations of the south-eastern half of 47\,Tuc (Fig.~\ref{fig:img})
as part of the {\it Spitzer} Surveying the Agents of Galaxy Evolution
Legacy program (SAGE-SMC). The SAGE-SMC observations consist of two
epochs, separated by three months, of Infrared Array Camera (IRAC)
maps at 3.6, 4.5, 5.8, and 8~\micron{}. Here, we use photometry
obtained by combining both epochs. Exposure times are short enough
(12~s) to ensure that bright sources remain unsaturated, but long
enough to obtain sufficient sensitivity for reasonably complete
photometry to beyond the HB \citep[cf.][]{boyer09a}. Angular
resolution with IRAC ranges from 1.7\arcsec{} at 3.6~\micron{} to
1.9\arcsec{} at 8~\micron{}.  The SAGE-SMC MIPS observations (24 --
160~\micron{}) did not cover 47\,Tuc. For details regarding data
acquisition and reduction for SAGE-SMC, see K.D.Gordon et al. (2010,
in preparation).

47\,Tuc was also observed with IRAC as part of PID20298
(P.I. R.Rood). We obtained these data from the {\it Spitzer} archive
and reduced them independently (hereafter referred to as the Rood
data). The Rood field-of-view at each IRAC wavelength reaches a cluster
radius of $\approx$10\arcmin{}, with an overlapping area of 5\arcmin{}
$\times$ 9\arcmin{} centered on 47\,Tuc (Fig.~\ref{fig:img}).  Two
depths were achieved: 12 s pix$^{-1}$ and 936 s pix$^{-1}$, hereafter
described as the shallow and deep data, respectively. Observation
details are described in \citet{origlia07}.  Here, these data were
processed with pipeline S18.7.0. The Basic Calibrated Data were
reduced and mosaicked with the MOPEX reduction package
\citep{makovoz05} after applying an array distortion correction. We
implemented the MOPEX overlap routine to match backgrounds between
overlapping areas of the images and the MOPEX mosaicker for outlier
elimination, image interpolation, and co-addition. The final mosaics
have pixel sizes of 0.86\arcsec{}~pix$^{-1}$.

\section{Photometry}
\label{sec:photometry}

K.D.Gordon et al. (2010, in preparation) contains details about
SAGE-SMC point-source photometry.

Point-source extraction ($> 4 \sigma$) was performed on the Rood data
with the DAOphot II package \citep{stetson87}, with point-spread
functions (PSFs) derived from $>$10 isolated stars in each mosaic.
Extended sources and outliers broader or narrower than the PSF were
rejected based on sharpness/roundness cut-offs.  The final fluxes are
color-corrected using a 5000~K blackbody \citep{handbook}, an
appropriate temperature for a typical RGB star. This correction
differs by $<$1\% from that for a 2000~K blackbody. A
pixel-phase-dependent correction \citep{reach05} was applied to the
3.6~\micron{} photometry.  Photometric errors include standard DAOphot
errors and the IRAC absolute calibration errors
\citep{reach05}. Magnitudes relative to $\alpha$\,Lyr (Vega) are
derived using the zero magnitudes from the {\it Spitzer} IRAC
data handbook \citep{handbook}. False star tests including both the
core and outskirts of the cluster indicate the photometry is
$\gtrsim$85\% complete at the HB in both the deep and shallow mosaics
at all four IRAC wavelengths. Bolometric corrections were determined
using 2MASS $J$- and $K_{\rm s}$-band photometry \citep{skrutskie06},
the transformation from \citet{montegriffo98}, and the reddening
($E(B-V) = 0.04~{\rm mag}$) and distance modulus ($(m-M)_0 =
13.32~{\rm mag}$) used in \citet{origlia07}.

\begin{figure}
\epsscale{1.1} \plotone{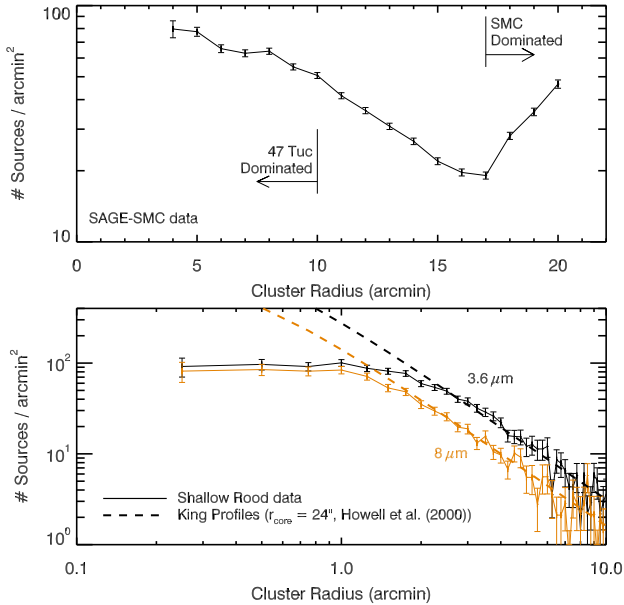} \figcaption{{\it Top:} Source density
profile in SAGE-SMC data. The SMC begins to dominate beyond $R \gtrsim
15\arcmin{}$. We restricted SAGE photometry to $R < 10\arcmin{}$ (see
text). {\it Bottom:} Source density profiles for the shallow Rood
data.  Dashed lines are King profiles with $R_{\rm core} = 24\arcsec$.
Note the data deviate from the King profiles due to source confusion,
inside 2.2\arcmin{} at 3.6~\micron{} and 1.8\arcmin{} at 8~\micron{}.
\label{fig:profile} }
\end{figure}

The shallow Rood data are confusion limited at $R \lesssim 2\arcmin$
(Fig.~\ref{fig:profile}). Our analysis suggests that the 3.6~\micron{}
data deviate from a King profile at $\approx$$2.2\arcmin$
\citep[$r_{\rm core}=24\arcsec$;][]{howell00}, while the 8~\micron{}
data deviate near 1.8\arcmin{}. Assuming the \citet{origlia07}
analysis is limited by the 8~\micron{} data (since they use
high-quality $K$-band data instead of 3.6~\micron{}), we might expect
to have found up to 25\% fewer sources in the inner 2\arcmin{} than
were detected in their analysis due to greater incompleteness at
3.6~\micron{} compared to 8~\micron{}.

The SAGE-SMC 3.6 and 8~\micron{} mosaics cover the south-eastern side
of the cluster with $3\arcmin \lesssim R < 10\arcmin$. We include only
sources within 10\arcmin{} to match the Rood coverage
(Fig.~\ref{fig:img}) and limit contamination from the SMC, which
begins to dominate the source density at $R \gtrsim$15\arcmin{}
(Fig.~\ref{fig:profile}). 

\begin{figure*}
\epsscale{1.15} \plotone{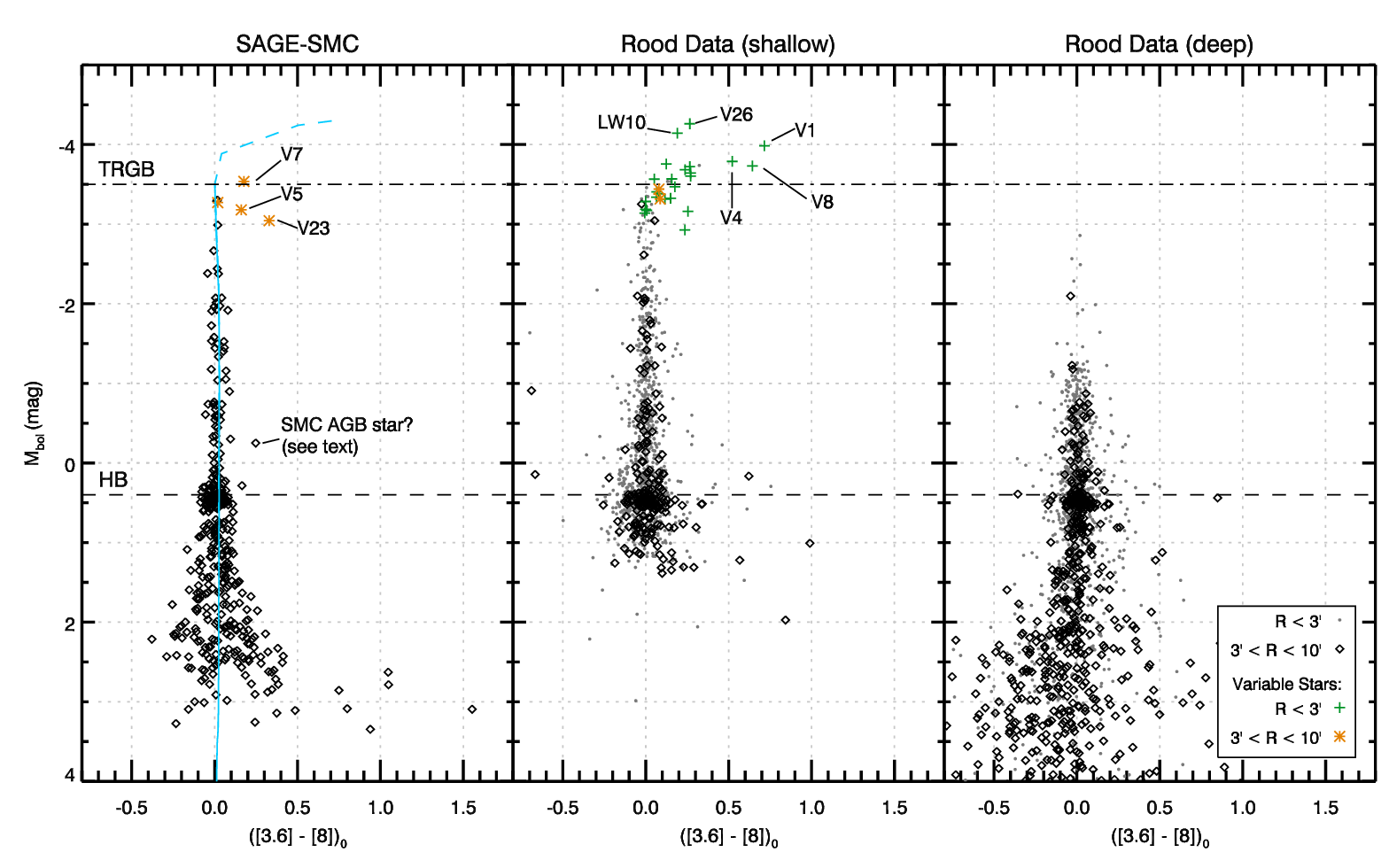} \figcaption{CMDs of 47~Tuc. {\it Left:}
SAGE-SMC, {\it Middle:} Shallow Rood data, {\it Right:} Deep Rood
data. Note that we see far fewer red sources here than seen by
\citet{origlia07} {\it also in the same original data}. The blue line
in the left panel is an isochrone from \citet{marigo08}, with $Z =
0.0024$ and $t_{\rm age} = 11.3$~Gyr; the dashed portion marks the
AGB. The black dot-dashed line marks the TRGB luminosity, as indicated
by the isochrone. Saturated sources are excluded from the deep
CMD. Variable stars LW10 and V26 may trace the AGB, while the
other variable stars trace the RGB.\label{fig:cmd} }
\end{figure*}

\section{Discussion}
\label{sec:disc}

\subsection{Dust Excess}
\label{sec:dust}

The CMDs in Figure~\ref{fig:cmd} include SAGE-SMC and Rood data.  RGB
star bolometric magnitudes peak in the near-IR, and warm circumstellar
dust is detectable as photometric excess at 8~\micron{}. The envelopes
we attempt to detect are optically thin at IR wavelengths (those that
are not would stand out); attenuation of starlight can thus be
neglected, whilst dust emission is insignificant at 3.6~\micron. Hence
it does not matter whether the $K-[8]$ color is used, as in
\citet{origlia07}, or the $[3.6]-[8]$ color. Indeed, in the re-reduced
Rood data, {\it all} stars with red $K_{\rm s}-[8]$ colors on the
upper half of the RGB (where photometric scatter is small) also have
red $[3.6]-[8]$ colors. We also note that the mean $K_{\rm
s}-[3.6]_{\rm SAGE}$ color for 47\,Tuc is a mere 0.08~mag,
insignificant compared to the range in dust excess ($0<K-[8]<1$)
suggested by \citet{origlia07}. We prefer to use the $[3.6]-[8]$ color
because it does not mix different origins of data, and we can
confidently apply the stellar blending test described in
Section~\ref{sec:blending}.

\subsection{SAGE-SMC CMD}
\label{sec:sage}

The SAGE-SMC CMD is presented in Figure ~\ref{fig:cmd}. The blue line
represents an isochrone from \citet{marigo08} with the appropriate age
(11.3~Gyr) and metallicity ($Z = 0.0024$) for 47\,Tuc from
\citet{thompson09}; the solid and dashed lines represent the RGB and
AGB, respectively. The isochrone indicates that the TRGB is located at
$M_{\rm bol} \approx -3.5~{\rm mag}$, agreeing with the luminosity
marking the onset of dust production in 47~Tuc \citep[$L \approx
2000~L_\odot$;][]{lebzelter06}.

Aside from 3 variable stars within 0.5~mag of the TRGB
\citep{lebzelter05}, only one source $\approx$1~mag brighter than the
HB is red, with $[3.6]-[8] \approx 0.25$~mag (R.A.$=00^{\rm h}25^{\rm
m}39\fs4$, decl.$=-72\degr08\arcmin21\farcs0$, J2000). This source is
not identified as variable by either \citet{weldrake04} or
\citet{lebzelter05}, but its 3.6~\micron{} magnitude varies by
0.14~mag between the two SAGE epochs. This source is isolated, and is
not affected by stellar blending or other imaging artifacts, so the IR
excess is likely real. However, the source is in the outskirts of the
cluster, in the direction of the SMC, and may not belong to
47\,Tuc. If at the distance of the SMC, its absolute 3.6~\micron{}
magnitude is $M_{3.6} = -7.75$~mag, which is $>1$~mag brighter than
the TRGB and consistent with a typical AGB star
\citep{bolatto07}. Thus we conclude that significant IR excess in
47\,Tuc is limited to $<1$~mag from the TRGB in the SAGE-SMC data,
similar to what is seen in $\omega$\,Cen, NGC~6791, and NGC~362
\citep{boyer08,vanloon08a,boyer09a}.

\begin{figure}
\epsscale{1.15} \plotone{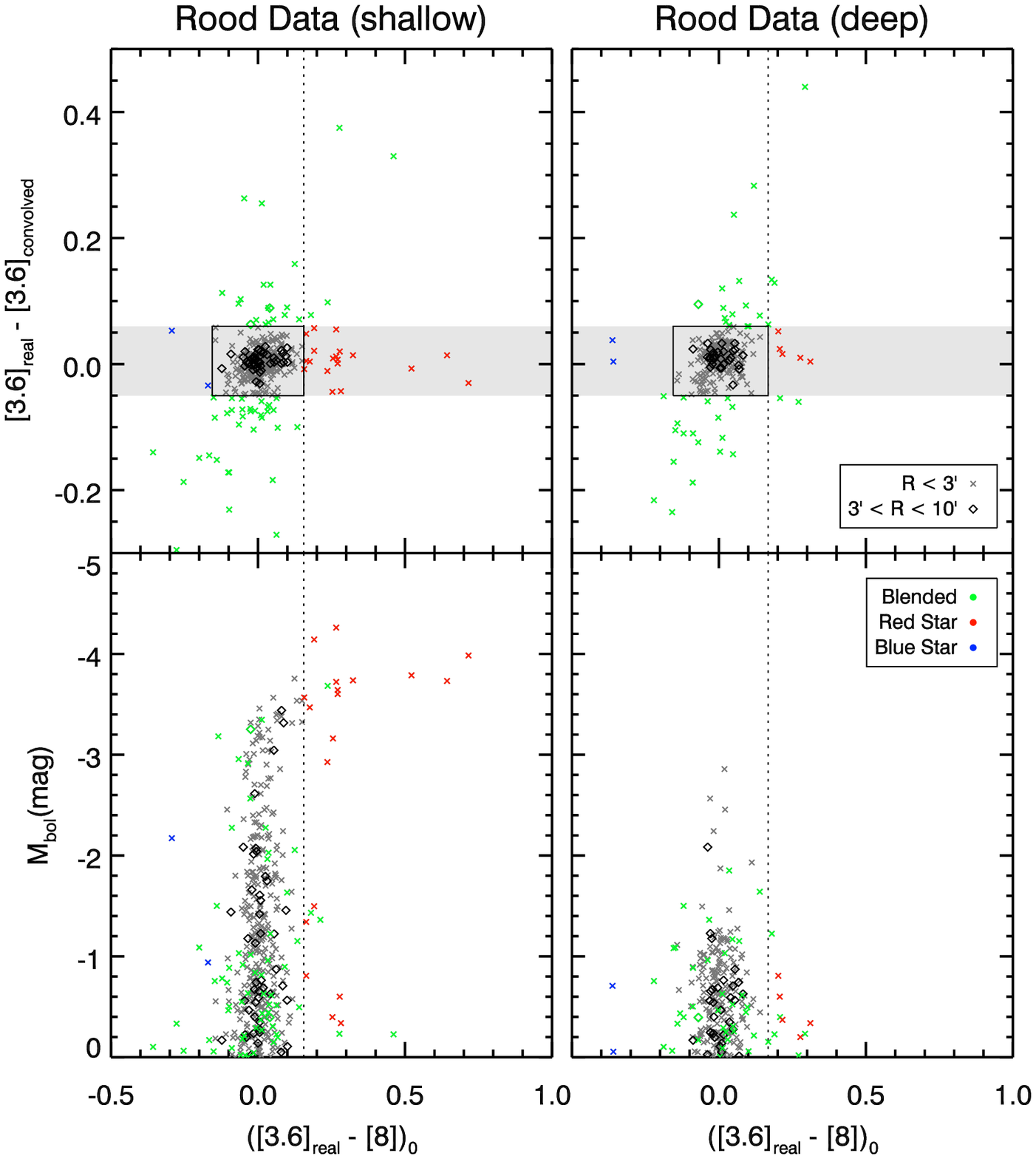} \figcaption{{\it Upper panels:} Source
  blending and confusion at 8~\micron{} above the HB.  See
  Section~\ref{sec:blending} for a description of the test. {\it Lower
  panels:} CMDs showing the locations of stellar blends, truly red
  sources, and truly blue sources.  The mean 3\,$\sigma$ excess
  ($[3.6]-[8] > 0.17$~mag) is marked by a dotted line. Note most red
  stars are confined to within 1~mag of the TRGB, and most stellar
  blending occurs among the faintest stars. \label{fig:blends} }
\end{figure}

\subsection{Spitzer Archive CMDs}
\label{sec:origlia}

Photometry extracted from the Rood data is noticeably noisier than the
SAGE-SMC photometry (Fig.~\ref{fig:cmd}, middle, right panels).  This
could partially be due to differences in data acquisition and
reduction, but it appears the scatter is caused mainly by inclusion of
the inner 3\arcmin{} of the cluster (light gray dots in
Fig.~\ref{fig:cmd}), where crowding is an issue. The deeper photometry
suffers from saturation at 3.6~\micron{} already more than a magnitude
{\it below} the TRGB. For clarity, saturated sources are excluded from
the CMD in Figure~\ref{fig:cmd}.

We see immediately that the new photometry has resulted far fewer
IR-excessive sources than found by \citet{origlia07}. That study found
$>90$ IR-excessive stars ($R<2\arcmin$) distributed evenly along the
entire RGB, while we see only a handful of red sources only near the
upper RGB, or in increasing number and redness towards magnitudes
fainter than $\approx$2~mag below the TRGB. In the shallow data
($R<10\arcmin$), we find 24 sources with $>3\,\sigma$ IR excess
($[3.6]-[8] \gtrsim 0.17$~mag) and $M_{\rm bol} < 0$~mag, four times
fewer than found by \citet{origlia07} in the same area of their
CMD. This is consistent with an AKARI analysis of 47\,Tuc presented by
\citet{ita07}, where it can be seen in their Fig.~2 that most dusty
stars reside near the TRGB.

Thirteen of these 24 red stars are variable stars identified by
\citet{lebzelter05}. The brightest variables, V26 and LW10, may be AGB
stars, whilst the others trace the RGB.  Curiously, both show no
excess in $8 - 13$~\micron{} spectra \citep{vanloon06}, but do show
moderate excess in the Rood data ($[3.6]-[8] \approx 0.2$~mag), which
were taken one month prior to the spectral observations. The
difference could be due to episodic dust production during the
pulsation cycle. The variables V1 and V8 are the reddest stars in the
shallow Rood data, agreeing with the dust excess seen in their mid-IR
spectra \citep{ramdani01,origlia02,lebzelter06,vanloon06}.

When excluding dusty variable stars, we find only 11 sources with
IR excess in the shallow Rood data, all but one more than
$\approx$2~mag fainter than the TRGB. These sources all fall within
3\arcmin{} of the cluster center, suggesting that stellar crowding may
be leading to inflated fluxes, as measured in the lower resolution
8~\micron{} images.

In the deep data, we find 11 other sources with $[3.6]-[8]$ excess
$>3\,\sigma$.  However, excluding the inner 2\arcmin{} of the cluster
as done in \citet{origlia07} leaves only one source with IR excess.

\subsection{Stellar Blending and Imaging Artifacts at 8~\micron{}}
\label{sec:blending}
A re-analysis of the photometry has revealed only 24 red sources,
missing 74\% of the 93 red sources from \citet{origlia07}.  This
discrepancy cannot be entirely due to the 25\% incompleteness of the
3.6~\micron{} data described in Section~\ref{sec:photometry} since
that would require red sources to be preferentially undetected
compared to non-red sources. That scenario is unlikely since the red
sources are evenly distributed in the cluster and since optically thin
circumstellar dust emission is minimal at 3.6~\micron{}. To investigate
whether the 24 red sources are truly harboring circumstellar dust, we
have performed a stellar blending test, similar to that done for
$\omega$\,Cen in \citet{boyer08}. We convolved the 3.6~\micron{} image
with the 8~\micron{} PSF and extracted point-source fluxes from the
resulting low-resolution 3.6~\micron{} image using the same procedure
described in Section~\ref{sec:photometry}. In the ideal scenario,
where all stars are isolated, $[3.6]_{\rm real} - [3.6]_{\rm
convolved}$ should always be near zero. Any sources that show
$[3.6]_{\rm real} - [3.6]_{\rm convolved} > 0$~mag are affected by
stellar blending, which inflates the convolved 3.6~\micron{}
flux. Stars that are truly red should have $[3.6]_{\rm real} - [8] >
0$~mag {\it and} $[3.6]_{\rm real} - [3.6]_{\rm convolved} \approx
0$~mag. Figure~\ref{fig:blends} shows the results of this test.

In the crowded environment of a GC, blending can be complicated. The
test described above may also result in blue colors, caused by stellar
confusion within compact groups of stars.  With many overlapping PSFs,
it is no surprise that flux measurements can be uncertain. We choose a
conservative cut-off of $\vert [3.6]_{\rm real} - [3.6]_{\rm
convolved}\vert > 0.05$~mag to define stars with affected photometry.

The boxes in Figure~\ref{fig:blends} show the mean 3\,$\sigma$
$[3.6]-[8]$ excess in the x-direction and the cut-off for blended
sources in the y-direction. The shaded regions to the left and right
of these boxes show the approximate locus of truly red or blue IR
colors. Stars above or below the shaded region are suffering from
blending and/or source confusion. Stars shown to have real red or blue
$[3.6]-[8]$ colors are marked in red and blue, respectively, and
sources affected by blending and/or confusion are marked in green. The
locations of these stars on the CMDs are shown in the lower panels. In
the shallow data, the horizontal plume of stars at the TRGB is
striking and very well distinguished from the vertical dispersion,
which is a result of stellar blending and source confusion.  Note that
in the shallow data, only one red star with $M_{\rm bol} \lesssim
-3$~mag is blended: variable LW8.

Of the 11 sources in the shallow data (identified in
Section~\ref{sec:origlia}) brighter than the HB and not identified as
variable stars, we find a total of 6 unblended, red sources below the
TRGB, all with $M_{\rm bol} \gtrsim -2$~mag. In the deep data, we find
5 such sources. Tellingly, none of these sources are red in {\it both}
the shallow and deep data.

Closer inspection of the shallow 8~\micron{} mosaic shows that all 6
of the faint, red, unblended stars fall directly on image artifacts
caused by internal optical scattering inside the detector array
\citep[known as banding;][]{handbook}, causing an artificial increase
in their 8~\micron{} fluxes. Therefore, there are no truly red sources
below the TRGB in the shallow data. Of the 5 faint, red, unblended
stars in the deep data, 2 are affected by banding, and 2 lie
immediately adjacent to a saturated star, which may affect both their
3.6 and 8~\micron{} fluxes. In the deep data, we therefore find only
one potentially dusty source below the TRGB. Again, we note that this
source is {\it not} red in the shallow data, indicating that the IR
excess is not real. Therefore, there are no truly dusty sources
fainter than $\approx$0.5~mag from the TRGB, consistent with the SAGE-SMC data.

\subsection{Implications for Mass Loss and Dust Production}
\label{sec:mdot_rgb}

To be consistent with the analysis from \citet{origlia07}, we would
have detected up to 70 dusty sources, even after considering
incompleteness in the 3.6~\micron{} data compared to their $K$-band
data. However, through close inspection of stellar blending and
imaging artifacts in SAGE-SMC data and the re-reduced
\citet{origlia07} data, we find that the number of truly dusty stars
in 47\,Tuc below about a magnitude fainter than the TRGB is consistent
with zero. We therefore find that the mass loss law derived by
\citet{origlia07} over-predicts mass loss along the RGB --
estimating nearly two orders of magnitude more mass loss among the
faintest RGB stars than Reimers' law \citep{reimers75} and attributing
$\approx$25\% of the total mass loss to stars fainter than $M_{\rm
bol} = -1.5$~mag. With no dusty mass loss along the RGB, these faint
RGB stars together contribute closer to 1\% of the total cluster mass
loss (based on Reimers' law). We do see a large number of red sources
at or above the TRGB (Fig.~\ref{fig:cmd}), as expected if the stellar
structure only becomes favorable to dust production if the star is
either close to the occurrence of core helium ignition (i.e., near the
TRGB), or is experiencing helium-shell flashes (i.e., on the thermally
pulsing AGB).

Very little, if any, dust is forming below the TRGB, but that is not
to say that no mass is lost on the RGB. Indeed, the morphology of the
HB requires that mass is lost on the RGB. Chromospherically-driven
winds are indicated in several RGB stars by asymmetries and coreshifts
in chromospheric spectral lines, with mass-loss rates up to $10^{-8}
M_\odot~{\rm yr}^{-1}$
\citep[e.g.,][]{mauas06,mcdonald07,dupree09}, which is almost
exactly what is required to explain the HB morphology.

In addition to only residing near the TRGB, the majority of the stars
showing evidence for circumstellar dust are also variable. This
supports the idea that dust production coincides with
pulsationally-enhanced stellar winds.

\acknowledgements We thank the anonymous referee for his/her thorough
review. This work was supported by {\it Spitzer} via JPL contracts
1309827 and 1340964 and makes use of data products from 2MASS, a joint
project of the University of Massachusetts and IPAC/Caltech, funded by
NASA and the NSF.


\clearpage

\end{document}